\documentstyle[prb,aps,psfig]{revtex}

\begin{document}

\draft 

\title{
Coherent electron-phonon coupling and polaron-like transport
in molecular wires}
 
\author{H.~Ness \cite{hnemail}, S.A.~Shevlin and 
A.J.~Fisher \cite{ajfemail}}

\address{Department of Physics and Astronomy,
University College London, Gower Street, \\
London WC1E 6BT, United Kingdom}

\maketitle

\begin{abstract}

We present a technique to calculate the transport properties through
one-dimensional models of molecular wires.
The calculations include inelastic electron scattering due to
electron-lattice interaction.
The coupling between the electron and the lattice is crucial to
determine the transport properties in one-dimensional systems subject
to Peierls transition since it drives the transition itself.
The electron-phonon coupling is treated as a quantum coherent process,
in the sense that no random dephasing due
to electron-phonon interactions is introduced in the scattering wave 
functions.
We show that charge carrier injection, even in the tunneling regime,
induces lattice distortions localized around the tunneling electron.
The transport in the molecular wire is due to polaron-like propagation.
We show typical examples of the lattice distortions induced by
charge injection into the wire. 
In the tunneling regime, the electron transmission is strongly enhanced 
in comparison with the case of elastic scattering through the undistorted 
molecular wire.
We also show that although lattice fluctuations modify the electron
transmission through the wire, the modifications are qualitatively
different from those obtained by the quantum electron-phonon
inelastic scattering technique.
Our results should hold in principle for other one-dimensional atomic-scale 
wires subject to Peierls transitions.

\end{abstract}



\section{Introduction}\label{intro}

Developments in nanofabrication, including both `bottom-up' approaches
and `top-down' methods are focussing renewed attention on the
properties of molecular-scale entities and their potential as
electronic device components \cite{wiresbook}.  
One of the most basic theoretical
questions that can be asked in this context is: what determines the
conductance of a molecule if it is used as a current-carrying element
bridging two reservoirs of differing electron chemical potential?
This question now has an immediate relevance for experiments in which
such conductances are measured, 
using either (i)
scanning probe tips for individual molecules adsorbed on surfaces
\cite{joachim95,joachim97,ad_on_surf,venema99}, 
for molecular wires adsorbed at step edges \cite{langlais99}, 
embedded in self-assembled monolayers \cite{bumm96,bumm99},
or (ii) (macroscopic) electrodes obtained by nanolithography
\cite{bockrath97,tans98,frank98,zhou00,porath00} or from a 
mechanically controllable break junction.
\cite{reed97,kergueris99a,kergueris99b}.

Since the seminal work of Aviram and Ratner \cite{aviram74} concerning
the electron transfer rate between acceptor and donor groups
linked by a conjugated molecular bridge, numerous theoretical studies
on electron transfer and transport through molecular systems have been
performed.  
In the following, we briefly review some contributions on the electron 
transport through a single organic molecule (or a few molecules) 
whose ends are connected to electron reservoirs.  
Calculations of the electronic transmission through such systems have been 
done for purely one-dimensional models
\cite{sautet88,joachim96,mujica96,magoga98,english98,onipko98-99,mujica00,hall00}
and two-dimensional models
\cite{nakanishi98-00,yaliraki98,onipko00}.  
More realistic descriptions of the electrode/molecule system have also
been developed.  
Combining elastic electron scattering theories with three-dimensional 
tight-binding-like Hamiltonians, models have been developed for molecular 
wires connected to two semi-infinite surfaces
\cite{kergueris99b,joachim96,joachim91,magoga97+99} or to two semi-infinite 
`rods' \cite{emberly98-99}, or to cluster-like
leads where imaginary parts are introduced in the Hamiltonian to take
into account the fact that electrons can leak into the metallic
reservoirs \cite{hall00,samanta96,datta_co97-99}.
Within a framework equivalent to the latter model, calculations have been
extended to the Hartree-Fock level for a molecule attached to gold
clusters \cite{yaliraki99}.  
More recently, density functional theory has been applied to a molecular 
wire (described by atomic pseudopotentials) connected to two jellium 
surfaces \cite{diventra00}.

These theoretical studies have clarified the importance of three major
points crucial for the transport properties of the molecular wire
connected to the reservoirs.  First, they have shown the importance of
the electronic and chemical interaction between the ends of the
molecular wire and the reservoirs.  The larger the Hamiltonian matrix
elements between the delocalized electronic states of the electron
reservoirs and those molecular electronic states that extend along the
wire, the better the conductance properties will be; furthermore, these
matrix elements should be large compared with the characteristic
intra-molecular Coulomb interaction between electrons in order to
avoid the Coulomb blockade \cite{C_blokade_book}.  Second, the
theories show that in the limit of small applied voltage and away from
the Coulomb blockade regime, the transport is dominated by charge
carrier tunneling inside the HOMO-LUMO gap of the molecule.  This gap
is another crucial parameter for control of the conductance of the
wire.  The smaller the gap is, the larger the tunneling transmission
will be.  More generally, the gap of a molecule depends on the
chemical nature and atomic structure of the system.  This gap can also
be modified by the electron-electron interactions or by a change of
the structure of the molecule due to ({\it i}) external forces, ({\it
ii}) (thermal) lattice fluctuations or ({\it iii}) electron-lattice
interaction.  Third, the calculations highlight the importance of the
position of the molecular electronic levels with respect to the Fermi
levels of the reservoirs in the presence of an applied voltage, and
the related issue of where the potential drops occur inside the
junction.  This has been done both empirically \cite{datta_co97-99}
and by using approximate \cite{mujica00} or exact \cite{diventra00}
self-consistent schemes.

Although the above theoretical work has shed light on several
important physical processes for the transport in molecular wires and
has matched to a certain level of accuracy with some experimental
measurements, all the models previously cited are based on elastic
electron scattering through the rigid lattice of the wire.  However,
in such highly confined electron systems, the coupling between
electron and other excitations (phonons for instance) is strongly
enhanced because of the size of the system and its quasi
one-dimensionality.  This makes the rigid lattice approximation
questionable---particularly so since a one-dimensional metal is
generically unstable to a Peierls transition at low temperature
\cite{peierls55}.  In an infinite system, such a transition typically
produces a semiconductor in which the states near the band extrema are
very strongly coupled to distortions of the system; in a conjugated
organic molecule, the corresponding phenomenon is a strong coupling of
the $\pi$-electrons occupying the HOMO and LUMO states to the
bond-alternation pattern.  This coupling means that the low-lying
states of a charged molecule (via which any net transport of charge
through the molecule must proceed) involve an intimate coupling of
electronic and lattice degrees of freedom, to produce excitations such
as polarons or solitons \cite{SSH88,lubook}.  
These coupled excitations can be thought of as conspiring to lower the 
energy gap locally around a charge carrier when it is introduced into the 
system.
Such polaronic and solitonic phenomena have been studied in bulk or thin 
film samples of conducting polymers for decades 
\cite{SSH88,bredas96+00,campbell97,conwell97}.

The importance of this electron-lattice coupling means that the
conventional manner of introducing lattice vibrations within a
Landauer-type approach to conductance, as an extra broadening of the
electronic levels (extra imaginary part in the corresponding
Hamiltonian, see for example Refs. \cite{datta_co97-99,dattabook}), 
is not sufficient to describe the
coherent lattice distortion due to charge injection. 
To our knowledge, the explicit nature of the distortion accompanying charge 
injection has only so far been partially addressed in two simplified 
limits \cite{note_tun_inel}.  
In the first case, a molecular wire was treated as a rigid lattice
in which a static soliton-like defect is present \cite{olson98}.  
Although it possesses a mid-gap electronic state,
this model does not permit the study of the dynamics of formation and
transport of charge-induced lattice distortions.
In the second case, the atoms of a conjugated molecule were assumed to
respond classically to the injection of a electron wave-packet
\cite{yu99}, 
via forces calculated from expectation values of the electron wave-packet 
and other electronic states.  
In this model, the lattice is able to respond to the injected
charge, but not in the physically correct manner: within a wave-packet
approach to tunneling, only a small part of the electronic charge
enters the tunnel barrier.  Therefore the lattice responds with
probability unity to a small fraction of the charge of the injected
particle, rather than responding with a small probability to the total
charge of the injected particle.

The only way to overcome these limitations is to perform transport
calculations in which the full dynamical correlation between charge carriers
and quantum phonons is retained.  
We report the results of such calculations in this paper.
In order to focus on this particular mechanism for charge transport, we use 
a simple tight-binding model of a conducting polymer 
(the Su-Schrieffer-Heeger (SSH) model for trans-polyacetylene \cite{SSH88}) 
that does not explicitly include any electron-electron interactions.
However, our calculations cover a range of transport regimes that includes 
tunneling transport (virtual electrons) and resonant transport (where there 
is sufficient energy to inject real electrons into the system).  
In contrast to the more usual `phase-breaking' approaches to the
electron-phonon interaction in transport problems, 
we explicitly retain the phase coherence between elastic and inelastic 
processes.
In this paper we concentrate on systems where the boundary conditions on the
molecular wires force them to be semiconducting in the zero-bias limit.  
Within the SSH model, this correponds to molecular chains containing an even 
number of monomers.
Calculations on odd-length chains, which incorporate mobile solitonic 
defects with associated mid-gap states that make an additional contribution 
to the transport, will be reported separately.

The paper is organized as follows. In Section \ref{model}, we present
the multichannel scattering technique used to calculate the transport
properties of the molecular wires.  
This section also includes a
detailed analysis of different approximations used for modelling the
molecular wires (involving harmonic phonons) and for reducing the
computational cost of the calculations (involving reduction of the
parameter space).  
The results obtained, in the limit of low temperatures, for the response 
of the molecular wires to charge
carrier injection are given in Section \ref{results}.  
We show how the lattice is distorted by the injection of a tunneling 
electron and how the coherent coupling between the tunneling electron and 
the quantum phonons affects the transmission properties through the wires.  
We also compare our transport results with the effect of straightforward 
static fluctuations in the harmonic lattice, excluding the dynamical 
correlation of the electrons and the phonons.  
Finally we summarize the most important results and propose further
developments of the present study (Section \ref{ccl}).  
Additionally, in an appendix we recall briefly the methods used to get 
the ground state and the harmonic phonon modes from the
original SSH model.  We also derive in the appendix the quantum
electron-phonon Hamiltonian used for the molecular wires.  
A brief account of part of this work has already appeared \cite{ness99}.

\section{Model}\label{model}

We are interested in modelling the coherent electron (or hole) transport 
through a finite size system (the molecular wire) connected to two leads 
which inject or collect the charge carriers.
Within the wire, the charge carriers interact with the atomic motion
which originally drives the Peierls transition in the molecule.

The interaction between the ends of the molecular wire and the 
leads is supposed to be strong enough to permit a good overlap between
the electronic states of the wire and the surface electronic wave functions
of the leads. 
In most of the practical applications, molecular wires end in ``active''
chemical groups, like thiol (S-H) for example, which are known to react easily
in presence of a gold surface to form chemical Au-S bonds \cite{AuSbond}.
We therefore assume that the electron transfer rate at the molecule/lead 
interface is such that we can consider the electron (hole) transport 
as being a coherent process throughout the nanojunction,  
rather than a sequential incoherent two-step process.

In the coherent transport regime, a stationary state wave function scattering
technique can be used to calculate the electron transfer through the molecular
wire.
Since we assume that the basis sets used to describe the leads and the 
molecular wire form a complete set, there are basically two ways to solve
the scattering problem for a single incident charge carrier.
The technique is reminiscent of the L\"owdin transformation \cite{lowdin62}. 
If one projects out the basis set associated to the molecular wire, the problem
is reformulated as a ``single impurity'' with on-site energies and coupling
matrix elements to the leads depending on the injection energy of the charge
carrier \cite{sautet88,magoga98}.
If one choose to project out the basis set associated to the leads, one can
effectively remove the leads from the problem. This technique is identical to 
the embedding technique where a (finite size) effective 
Hamiltonian describing the region of interest is obtained by introducing complex
embedding potentials \cite{dattabook,williams82,inglesfield81}. 
The embedding potentials characterise the matching of electronic spectrum of 
the wire to the continuum of states of the semi-infinite leads. 
These potentials also depend on the charge injection energy.

As we wish to obtain the response of the molecular wire to charge injection
as well as the transport properties through the junction, we choose the
embedding approach to solve the electron (hole) transport in the system.
For this, we use a technique which permits us to map the many-body 
electron/phonon problem onto a single-particle problem with many 
channels \cite{bonca95,anda94}.

In the remain of this section, we present the basis of the many-channel
scattering technique and discuss the different approximations introduced 
to reduce the computational cost of the calculations.  
Details of the construction of the quantum Hamiltonian for the molecular
wire from the model originally proposed by Su, Schrieffer
and Heeger (SSH) \cite{SSH88,SSH} are given in Appendix \ref{appA}.

Starting from this model, we have derived a quantum electron-phonon
Hamiltonian:
\begin{equation}
H_{\rm w}=
\sum_n \epsilon_n \ c^\dag_n c_n
+\sum_q \hbar\omega_q \ a^\dag_q a_q
+\sum_{q,n,m}\gamma_{qnm}(a^\dag_q+a_q)\ c^\dag_n c_m \ ,
\label{Hwire}
\end{equation}
where $c^\dag_n$ creates an electron in the $n$th one-electron state of the
molecular wire with energy $\epsilon_n$ and $a^\dag_q$ creates an excitation 
in the $q$th eigenmode of vibration (phonon) of the molecule with energy 
$\hbar\omega_q$.
The Hamiltonian Eq. (\ref{Hwire}) goes beyond the Holstein and Fr\"ohlich 
model for the electron-phonon ($e$-ph) interaction,
in the sense that the electron couples to different non-local eigenmodes of
vibration, each mode having a different frequency.
The electron-phonon coupling is linear in the phonon field displacement
and involves electronic transitions via a general form for the $e$-ph 
coupling matrix elements $\gamma_{qnm}$.

The electronic eigenstates and eigenvalues are determined self-consistently
with the atomic configuration for the ground state of the neutral dimerized
molecular chain taken to be the reference system (App. \ref{appA}).
From the atomic and electronic structures of the reference system, the phonon
modes and frequencies are calculated within the harmonic approximation 
(App. \ref{appA}).
The $e$-ph matrix elements are derived from the SSH model by expanding the
atomic displacements induced by adding a charge onto the phonon modes of 
the neutral molecule (App. \ref{appA}).
We have checked the accurary and the validity of the harmonic approximation
for the phonons (see below section \ref{approx}).

\subsection{Multichannel scattering technique}\label{MCST}

We are interested in solving the problem of electron transport through
a `nanojunction' within a two-terminal device.
We are mostly interested in the coherent regime for the electron transport,
that is the regime where no random phase breaking is arbitrarily introduced
between different electron scattering states.
Furthermore, we wish to use a formalism that can treat 
({\it i}) different transport regimes (pure tunneling, resonant tunneling, 
eventually ballistic transport) on an equal footing or in a transparent way, 
and ({\it ii}) the coupling of an electron with other degrees of freedom 
within the `nanojunction'.
In principle, to study the electronic transport in such open systems, one 
would have to deal with the non-equilibrium Green's functions formalism 
\cite{caroli71,wingreen88,meir92,anda91}.

In this paper, we use a model which maps a many-body problem (to be
accurate a one electron/many bosons problem) onto a single-particle problem
with many channels \cite{bonca95,anda94}. 
In a such model, one deals directly with the multichannel scattering states,  
although the formalism can be reformulated in terms of Green's functions.
This multichannel scattering technique has already been used to study
electron transport through one-dimensional models of ({\it i}) double-barrier 
resonant tunneling junctions with electron coupled to localised single 
phonon mode \cite{anda94,orellana96,makler00}, ({\it ii}) Holstein phonon 
model in presence of an electric field \cite{bonca97}, ({\it iii}) mesoscopic 
structures \cite{bonca95} and Aharonov-Bohm rings with on site phonon 
coupling \cite{bonca95,haule99}, ({\it iv}) tunneling barriers with the 
electron coupled to surface plasmon modes \cite{ness98}.
More recently such a technique has also been used to study inelastic electron 
tunneling through small molecules in a STM tunneling barrier 
\cite{mingo99,mingo00}.

We start with the following heterojunction: a molecular wire containing
$N_a$ atomic sites, described by the Hamiltonian $H_{\rm w}$ in 
Eq. (\ref{Hwire})
is connected to ideal one-dimensional right (R) and left (L) metallic leads, 
whose Hamiltonians are
\begin{equation}
H_{\rm R}=\sum_{l=N_a+1}^{+\infty} \epsilon_{\rm R}\ d^\dag_l d_l
+ \beta_{\rm R}(d^\dag_l d_{l-1}+d^\dag_{l-1} d_l)
\label{HR}
\end{equation}
and
\begin{equation}
H_{\rm L}=\sum_{l=-\infty}^{0} \epsilon_{\rm L}\ d^\dag_l d_l
+ \beta_{\rm L}(d^\dag_l d_{l-1}+d^\dag_{l-1} d_l) \ ,
\label{HL}
\end{equation}
via the coupling matrices,
\begin{equation}
T_{\rm R}=v_{\rm R} (d^\dag_{N_a+1} c_{N_a} + c^\dag_{N_a} d_{N_a+1})
\label{TR}
\end{equation}
and
\begin{equation}
T_{\rm L}=v_{\rm L}(d^\dag_0 c_1+c^\dag_1 d_0) \ .
\label{TL}
\end{equation}
The operators $d^\dag_l$ ($d_l$) create (annihilate) an electron on
site $l$ inside the leads, with on-site energy $\epsilon_{\rm L,R}$
and nearest neighbours hopping integrals $\beta_{\rm L,R}$, the
operators $c^\dag_i$ ($c_i$) are the corresponding operators for an
electron on site $i$ within the molecule, and $v_{\rm R}$ and $v_{\rm L}$ 
are the hopping matrix elements between the ends of the molecule and
the right and left leads respectively.  Within the molecular wire,
the transformation from site representation to eigenstate
representation is easily performed knowing that 
$c^\dag_n=\sum_{i=1}^{N_a} Z_n^*(i)\ c^\dag_i$, where $Z_n(i)$ are the 
components of the $n$th
electronic eigenstate of the wire (see Appendix \ref{appA}).

The procedure for mapping the problem onto a single-particle system 
with many channels is performed by writing the total scattering wave function
$\vert\Psi(E)\rangle$, for the total energy $E$ of the electron-phonon 
system as
\begin{equation}
\vert\Psi(E)\rangle=\sum_l\sum_{\{n_q\}}
\alpha_{l,\{n_q\}}(E)\ \vert l,\{n_q\}\rangle \ ,
\label{totwf}
\end{equation}
where the basis set used to expand the scattering waves is defined (in
the case of electron transport) as $\vert
l,\{n_q\}\rangle=c^\dag_l\prod_q \frac{(a^\dag_q)^{n_q}}{\sqrt{n_q!}}
\vert 0\rangle$ (for $1\le l\le N_a$) and $\vert
l,\{n_q\}\rangle=d^\dag_l\prod_q \frac{(a^\dag_q)^{n_q}}{\sqrt{n_q!}}
\vert 0\rangle$ (for other values of $l$), $\vert 0\rangle$ being the
vacuum state and $\{n_q\}$ being the phonon occupations.  The vacuum
state $\vert 0\rangle$ is taken to be the neutral ground state of the
system, with a definite number of electrons in each of the left lead,
right lead and molecule.  The electronic states we consider therefore
involve adding a single electron to this neutral state; the added
electron may be anywhere in the system (in the left lead, the right
lead, or the molecule).  For hole transport we use an identical basis,
except that electron creation operators are replaced by annihilation
operators.  By writing the wave function coefficients as in
Eq.(\ref{totwf}), no explicit separation between the electronic and
phonon degrees of freedom has been assumed.

As far as the electron (hole) propagation is concerned, each different 
channel is associated with a different set of phonon 
occupation numbers $\{n_q\}$.
The total wave function $\vert\Psi(E)\rangle$ is the eigenstate of the total 
Hamiltonian
$H=H_{\rm L}+T_{\rm L}+H_{\rm w}+T_{\rm R}+H_{\rm R}$, 
$H\vert\Psi(E)\rangle=E\vert\Psi(E)\rangle$, with the full scattering
boundary conditions applied ({\it i.e.} an incident electron or hole 
with the molecule in a given vibrational state).
 
In absence of dissipation, the total energy $E$ of the system is conserved
during the scattering process, {\it i.e}
\begin{equation}
E=E_{\rm in}+\sum_q n_q\ \hbar\omega_q
=E_{\rm out}+\sum_q m_q\ \hbar\omega_q \ ,
\label{Etot}
\end{equation}
where $E_{\rm in}$ is the energy of the incoming electron
and $\{n_q\}$ is the initial set of phonon occupation numbers.
$E_{\rm out}$ is the energy of the outgoing reflected or transmitted electron
with the corresponding set of phonon occupancies $\{m_q\}$.
For inelastic scattering processes ${\{n_q\}}\neq{\{m_q\}}$; for elastic
scattering the phonon distribution is conserved.

\begin{figure}
\psfig{figure=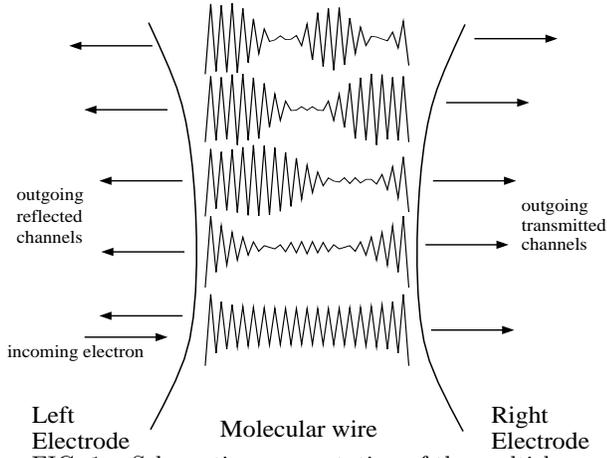,width=8cm,height=6cm}
\caption{
\label{fig_MCST}
Schematic representation of the multichannel configurations for
a molecular wire connected to two electrodes.
The different diagrams represent the bond length alternation
in the wire for the different channels. Initially the wire
is its ground state phonon configuration (lower diagram showing
a perfect bond length alternation in the middle of the wire).
The incoming electron can exchange energy with the phonon modes
inside the wire and therefore modify the initial bond length
pattern (other diagrams).}
\end{figure}

The wave function 
coefficients $\alpha_{l,\{m_q\}}$ take an asympotic form inside the leads. 
The form corresponds to propagating Bloch waves inside the left and right leads 
whose amplitudes are the reflection $r_{\{m_q\}}$ and transmission $t_{\{m_q\}}$ 
coefficients of the electron in the different channels.
For an injected electron from the left lead (for example), we have
\begin{eqnarray}
\alpha_{l,\{m_q\}}=
{e}^{{\rm i}k_{\{n_q\}}^{\rm L} l}\ \delta_{\{n_q\},\{m_q\}}
+r_{\{m_q\}}\ {e}^{-{\rm i}k_{\{m_q\}}^{\rm L} l}
\label{left_wf}
\end{eqnarray}
inside the left lead ($l\leq-1$) and
\begin{eqnarray}
\alpha_{l,\{m_q\}}=
t_{\{m_q\}}\ {\rm e}^{{\rm i}k_{\{m_q\}}^{\rm R} l}
\label{right_wf}
\end{eqnarray}
inside the right lead ($l\geq N_a+2$).  The $k_{\{n_q\}}^{\rm L,R}$
are the dimensionless wave-vectors of the Bloch waves in the different
channels.  For a given total energy $E$, the wave-vectors depend on
the phonon occupation numbers.  Figure \ref{fig_MCST} shows a
simplified sketch of the multichannel technique for different phonon
excitations inside the molecular wire.

The solutions of the Schr\"orindger equation 
$\langle l\vert H\vert\Psi(E)\rangle=E\langle l\vert\Psi(E)\rangle$ 
inside the leads give the dispersion relations for the electron wave-vectors 
inside the different channels,
$E_{\rm in}=\epsilon_{\rm L}+2\beta_{\rm L}\cos k_{\{n_q\}}^{\rm L}$ for 
the incoming wave and 
$E_{\rm out}=\epsilon_{\rm L}+2\beta_{\rm L}\cos k_{\{m_q\}}^{\rm L}$ for 
the outgoing reflected wave and to
$E_{\rm out}=\epsilon_{\rm R}+2\beta_{\rm R}\cos k_{\{m_q\}}^{\rm R}$ for 
the transmitted wave to the right lead.

In this paper, we report calculations valid in the limit of low
temperatures assuming that $k_{\rm B}T\ll\hbar\omega_q$. 
Therefore we take for the initial set of phonon occupation numbers, the set
corresponding to all phonon modes in the ground state $\{n_q\}=\{0\}$.
The dispersion relations become simply
$E=\epsilon_{\rm L,R}+2\beta_{\rm L,R}\cos k_{\{0\}}^{\rm L,R}$ for the
elastic channels and
$E=\epsilon_{\rm L,R}+2\beta_{\rm L,R}\cos k_{\{m_q\}}^{\rm L,R}+\sum_q
m_q\hbar\omega_q$ for the inelastic channels.
Note that in this case, the total energy $E$ also represents the electron
injection energy. 
 
Solving 
$\langle l\vert H\vert\Psi(E)\rangle=E\langle l\vert\Psi(E)\rangle$ 
at the interfaces of the heterojunction is the next step to perform
in order to resolve the unknown wave function coefficients inside 
the molecular wire.
For $l=0$ and $l=N_a+1$, the relations between the reflection
and transmission coefficients with the wave function coefficients at the
ends of the wire are obtained:
\begin{equation}
t_{\{m_q\}}=\frac{v_{\rm R}}{\beta_{\rm R}}\ \alpha_{N_a,\{m_q\}}\
{e}^{-{\rm i}k_{\{m_q\}}^{\rm R}N_a}\ ,
\label{t_coef}
\end{equation}
and 
\begin{equation}
r_{\{m_q\}}=-\delta_{\{0\},\{m_q\}}\ 
+\frac{v_{\rm L}}{\beta_{\rm L}}\ \alpha_{1,\{m_q\}}\ , 
\label{r_coef}
\end{equation}
for the elastic $\{0\}$ and inelastic $\{m_q\}\neq\{0\}$ channels.

Finally, solving the Schr\"odinger equation on sites $l\in[1,N_a]$
permits one to effectively remove the leads by introducing complex
embedding potentials \cite{dattabook}. 
Then
the solution of the full scattering problem is obtained by solving
the following complex linear system
\begin{equation}
\left[
E-H_{\rm w}-\Sigma^{\rm L}(E)-\Sigma^{\rm R}(E)
\right]
\vert\alpha(E)\rangle
=
\vert s(E)\rangle \ ,
\label{linsys}
\end{equation}
where $H_{\rm w}$ is the molecular wire Hamiltonian Eq. (\ref{Hwire}),
the components of $\vert\alpha\rangle$ are the scattering wave
function coefficients inside the molecular wire expressed in the
molecule eigenstate representation, {\it i.e.}
$\alpha_{n,\{n_q\}}=\sum_i Z_n(i)\ \alpha_{i,\{n_q\}}$,
and 
$\Sigma^{\rm L,R}$ are the embedding potentials due to the left and
right lead respectively.
The embedding potentials are diagonal matrices in the 
$\vert n,\{n_q\}\rangle$ basis set with components
\begin{equation}
\Sigma^{\rm L}_{n,\{n_q\}}(E)=
Z_n(1)\ v_{\rm L}\ g^{\rm L}_{\{n_q\}}(E)\ v_{\rm L}\ Z_n(1) \ ,
\label{SigmaL}
\end{equation}
and
\begin{equation}
\Sigma^{\rm R}_{n,\{n_q\}}(E)=
Z_n(N_a)\ v_{\rm R}\ g^{\rm R}_{\{n_q\}}(E)\ v_{\rm R}\ Z_n(N_a) \ ,
\label{SigmaR}
\end{equation}
where $g^{\rm L,R}_{\{n_q\}}$ are the surface Green's function of the 
isolated left and right leads for the different channels given by
\begin{equation}
g^{\rm L,R}_{\{n_q\}}(E)=\exp({\rm i}k_{\{n_q\}}^{\rm L,R}(E))
/\beta_{\rm L,R} \ .
\label{surfGfnc}
\end{equation}
Finally $\vert s(E)\rangle$ represents the source term ({\it i.e.} the
injected electron or hole at energy $E$) with components given by
\begin{equation}
s_{n,\{n_q\}}(E)=\delta_{\{0\},\{n_q\}}\ 
(-2{\rm i}\ v_{\rm L}\sin k_{\{0\}}^{\rm L})\ 
Z_n(1) \ .
\label{source}
\end{equation}
In the present case, the boundary condition chosen for the source term
corresponds to the injection of the charge, into the elastic channel
$\{0\}$, from the left lead towards the right lead.

The solution of Eq. (\ref{linsys}) can be obtained by several means using
algorithms for sparse matrices. 
In the present work, we explicitly separate the real and imaginary 
parts of vectors and matrices as follows
\begin{eqnarray}
\left[ 
\begin{array}{cc}
E-H_{\rm w}-\Re e\ \Sigma(E) & \Im m\ \Sigma(E) \\
\Im m\ \Sigma(E)               & -E+H_{\rm w}+\Re e\ \Sigma(E) \\
\end{array}
\right]
\left[
\begin{array}{c}
\Re e\ \alpha(E) \\
\Im m\ \alpha(E)
\end{array}
\right]
=
\left[
\begin{array}{c}
\Re e\ s(E) \\
-\Im m\ s(E)
\end{array}
\right] \ ,
\label{Clinsys}
\end{eqnarray}
where $\Re e$ and $\Im m$ denote the real and imaginary parts of the different
quantitites and $\Sigma=\Sigma^{\rm L}+\Sigma^{\rm R}$.
The matrix in the left-hand-side of Eq. (\ref{Clinsys}) is therefore real
and symmetric. We then use standard conjugate gradient (CG) technique to solve 
the linear system $A\vert x\rangle=\vert b\rangle$ where $A$ is a
real and symmetric matrix \cite{golubbook}. 
An efficient algorithm has been devised for computing the products 
$H_{\rm w}\vert x_i\rangle$ generated during the iterative CG steps.
It is based on an optimal adressing of the vector components and uses the 
selection rules for the $\gamma_{qnm}$ matrix elements 
(see Appendix \ref{QM_ph}).

\subsection{Isolated molecular wire : harmonic phonons and reduced parameter space}
\label{approx}

We have shown that the solution of the scattering problem is obtained by solving
Eq. (\ref{linsys}) for the value of the wave functions inside the molecular 
wire. 
Assuming a truncated phonon space up to a finite number of excitations 
$n_{\rm occ}^{\rm max}$, the size of the basis set is given by 
$N_{\rm size}=N_e\times(n_{\rm occ}^{\rm max}+1)^{N_{\rm ph}}$ 
with $N_e$ ($N_{\rm ph}$) being the number of electronic states
(phonon modes).
Even for relatively short wires,
the size $N_{\rm size}$ of the basis set quickly becomes too large for 
tractable numerical calculations and/or reasonable computing times.  
For example for $N_a=20$ atomic sites ($N_{\rm ph}=19$ acoustic and optic
phonon modes) with only $n_{\rm occ}^{\rm max}=2$,
we obtain $N_{\rm size}\gg 10^6$.
In the following we show how to reduce the basis set size and prove 
the validity of the approximations introduced.
The corresponding Hilbert space can be reduced by ({\it i}) considering
only valence (conduction) band electronic states (this will correspond
to hole (electron) transport respectively), by ({\it ii}) considering a 
limited but sufficient set of phonon modes and finally by ({\it iii}) using
only a few excitations in each phonon modes (truncated harmonic oscillator
approximation).

\begin{figure}
\psfig{figure=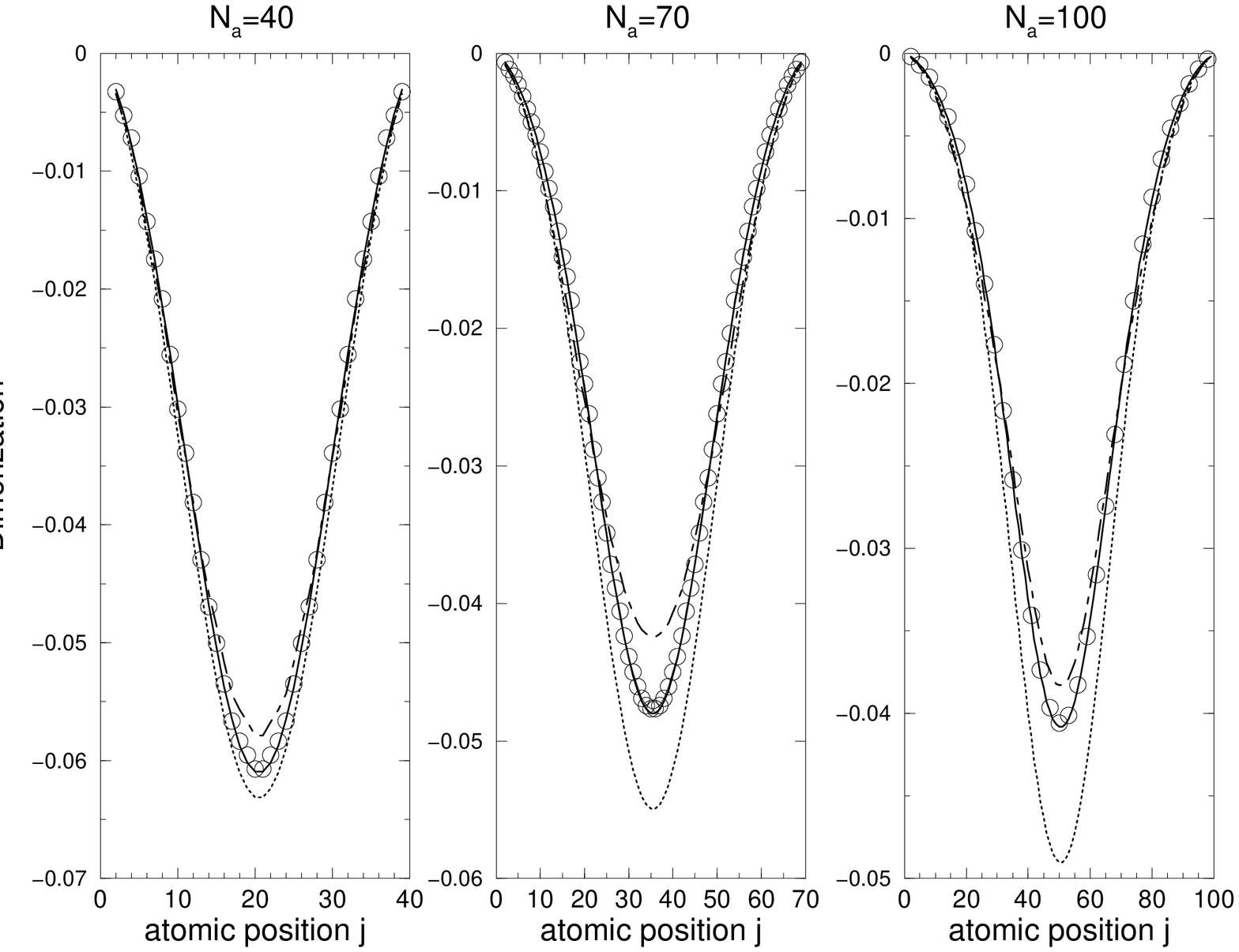,width=10cm,height=6cm}
\caption{
\label{static_disto}
Dimerization pattern $d_j$ (in \AA) induced by adding an extra electron 
into the molecular wire for three different wire lengths $N_a=40, 70$ 
and 100.
The dip in the dimerization patterns is characteristic of the formation
of a static polaron located in the middle of the chain.
The dotted lines correspond to the dimerization obtained form the original
SSH model with the full electronic spectrum. 
The dimerization obtained by considering half the electronic spectrum and
classical phonons is given by the solid lines (all the phonon modes), and
circles (6 optical modes for $N_a=40$ and 70 and 10 modes for $N_a=100$).
The dimerization calculated for quantum phonons is represented by
the dot-dashed lines (6 optical modes and $n_{\rm occ}^{\rm max}=6$ 
for $N_a=40$, 6 modes and $n_{\rm occ}^{\rm max}=4$ for $N_a=70$,
4 optical modes and $n_{\rm occ}^{\rm max}=5$ for $N_a=100$).}
\end{figure}

In order to determine which phonon modes are mainly contributing to the
charge-induced deformation of the chain, we calculate the ground state
atomic configurations for a neutral chain $u_i^0$ and for a chain charged
with one additional electron $u_i^{\rm c}$.
Note that because of the charge-conjugation symmetry (which holds exactly for
the SSH model) adding an extra electron or removing an electron ({\it i.e.} 
adding a hole) produces exactly the same lattice distortion.
Such a lattice distortion is known as an electron (or hole) polaron
({\it c.f.} Fig. \ref{static_disto}).
The lattice distortion is then projected onto the harmonic eigenmodes of 
vibration $V_q$ of the neutral chain and the corresponding Huang-Rhys 
factors $S_q$ are determined by \cite{huangrhys}
\begin{equation}
S_q=\frac{\frac{1}{2}M\omega_q^2\Delta_q^2}{\hbar\omega_q} \ ,
\end{equation}
where $\Delta_q=\sum_i V_q(i)\ (u_i^{\rm c}-u_i^0)$.
The Huang-Rhys factors give the averaged number of quantum phonons
that would be needed to achieve the corresponding elastic energy
of the lattice distortion.
We have checked that the largest values for those factors are obtained
for the lowest energy (longest wave-length) optical phonon modes
of the molecular wires \cite{note_ya-mode}. 
In particular, the optical phonon mode having the lowest energy
also has the most important contribution ({\it i.e.} $S_q>0.8$
as can be seen for different chain lengths in Table \ref{SQandEner}).

\begin{figure}
\psfig{figure=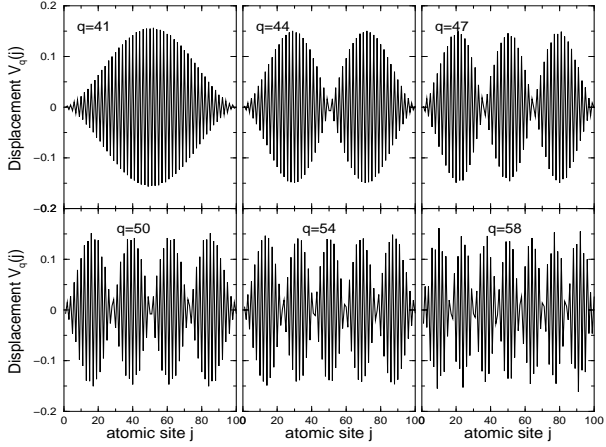,width=8cm,height=6cm}
\caption{
\label{fig_optmodes}
Lowest energy (longest wave-length) optical phonon modes $V_q(j)$
for the neutral chain containing $N_a=100$ atomic sites. The
frequencies (energies) of the modes are $\hbar\omega_q$=0.136, 0.141,
0.147, 0.153, 0.158 and 0.163 eV for the mode $q$=41, 44, 47,
50, 54 and 58 respectively. The phonon mode indexes $q$ of the
finite length chain are taken such that the (acoustic and optic)
phonons are ordered by increasing frequency.}
\end{figure}

From these results, we can already reduce the number of phonon modes
necessary by considering only the longest wave-length optical modes
to describe the lattice deformation induced by adding a charge into 
the chain.
As typical set of these modes is shown on Figure \ref{fig_optmodes}.

Furthermore, we can also check the validity of the harmonic approximation
used to determine the phonon modes of the wire.
Note that in section \ref{harmonic_ph}, the elastic energy is expanded
up to second order for small displacements around the equilibrium 
atomic positions in order to obtain the dynamical matrix from which the
eigenmodes of vibration are determined.
The deformation of the lattice due to charge addition is not purely
harmonic because of the distance dependence in the electron hopping
matrix elements, cf. Eq. (\ref{SSHamilt}).
However there are good reasons to believe that the potential surface
can be fairly well described by the harmonic approximation to the
neutral chain. 
To confirm this, we compare the deformation energy $\Delta E$ with
the harmonic distortion energy 
$E_{\rm harm}=\sum_q \frac{1}{2}M\omega_q^2\Delta_q^2
=\sum_q \hbar\omega_q\ S_q$.
The deformation energy is obtained from 
$\Delta E=E_0[u_i^{\rm c}]-E_0[u_i^0]$ where
$E_0[u_i^0]$ is the selfconsistent total energy of the SSH Hamiltonian
for a neutral chain with the corresponding atomic positions $u_i^0$.
$E_0[u_i^{\rm c}]$ is the non-selfconsistent total energy of the
neutral chain where the atomic positions $u_i^{\rm c}$ are taken to be 
those of the charged chain. 
In $E_0[u_i^{\rm c}]$ the effects of the imposed lattice distortions
on the electronic spectrum are taken into account.
Typical values of $\Delta E$ and $E_{\rm harm}$ are given in Table
\ref{SQandEner} for different wire lengths.
For short chains, the values of $\Delta E$ and $E_{\rm harm}$ are
almost indentical (less than $\approx$ 5\% difference).
For longer chains, the difference between $\Delta E$ and $E_{\rm harm}$
increases but never exceed $\approx$ 15\%.
We also show below that the harmonic expansion of the elastic energy
is sufficient to describe the formation of a static polaron by adding
a permanent extra charge in the molecular chains.

Now we turn on the reduction of the Hilbert space in relation to the
electron states. 
We want to check the validity of using only half the electronic spectum
on the values of the atomic displacements induced by adding a charge 
(electron or hole) into the molecular chain.
As mentioned in Ref. \cite{ness99}, we consider the action of the
model Hamiltonian Eq. (\ref{Hwire}) on the ($N_a\pm 1$)-electron
Hilbert space obtained by adding a charge into the neutral chain.
It is thought that it is sufficient to consider only the 
($N_a\pm 1$)-electron because for molecular wires strongly (electronically)
coupled to the leads, the mean time between charge passages is
$\approx 10^{-7}$ s (for a corresponding current of 1 pA), a value orders
of magnitude bigger than a typical residence time ($\approx 10^{-15}$ s).
Then we project out the addition of a charge into the electronic states
by working with the electronic eigenstates representation,
{\it i.e.} the sums over the eigenstates will only include the
occupied valence band states when adding a hole (and later for hole
transport) and only the empty conduction band states when adding an
electron (for electron transport).
In the following, we show that the static lattice distortions 
due to adding a charge are well reproduced by considering only one half
of the electronic spectrum.

We start from the ground state of the reference system (neutral chain with
atomic positions $u_i^0$).
The lattice distortions due to charging are expanded onto the harmonic
phonon modes $V_q(i)$ as $u_i=u_i^0+\sum_q V_q(i)\Delta_q$.
Then the influence of the lattice distortions on the
electronic Hamiltonian $H^0\equiv\epsilon_n\delta_{nm}$ are taken into
account by introducing the corresponding
$e$-ph coupling off-diagonal elements $\gamma_{qnm}$ in $H^0$.
The electronic Hamiltonian of the isolated molecule is
\begin{equation}
H^{\rm el}_{nm}=H^0_{nm}+H^{e\rm{-ph}}_{nm}=
\epsilon_n\delta_{nm}+\sum_q \tilde{\Delta}_q \gamma_{qnm}\ \ ,
\label{Hclass}
\end{equation}
where the matrix elements $\gamma_{qnm}$ are given 
by Eq. (\ref{gamqnm}) in Appendix \ref{appA},
and $\tilde{\Delta}_q$ is the dimensionless displacement
${\Delta}_q=\tilde{\Delta}_q\sqrt{\hbar/(2M\omega_q)}$.
The total energy of the distorted lattice is
\begin{equation}
E_{\rm half}(\{\Delta_q\})=
\sum_q \frac{1}{2}M\omega_q^2\Delta_q^2
+ {\rm Tr}[\rho^{\rm el}H^{\rm el}(\{\Delta_q\})] \ ,
\label{Eclassic}
\end{equation}
where $\rho^{\rm el}$ is the electronic density operator. 
To find the corresponding ground state, $E_{\rm half}(\{\Delta_q\})$ has to 
be minimised versus the classical lattice phonon displacements 
$\{\Delta_q\}$.

Within the half spectrum approximation, the trace in
Eq. (\ref{Eclassic}) runs only over the conduction (or valence) band
eigenstates when we consider the wire charged by an extra electron (or
hole).  Adding a charge to the system involves only the LUMO or HOMO
electronic state for an electron or hole respectively, therefore the
functional to be minimized is actually $E_{\rm half}(\{\Delta_q\})=
\sum_q \frac{1}{2}M\omega_q^2\Delta_q^2 + \lambda_e(\{\Delta_q\})$
where $\lambda_e=\langle\varphi_e\vert H^{\rm
el}(\{\Delta_q\})\vert\varphi_e\rangle$ is the lowest (or highest)
energy eigenvalue of the electronic half-space Hamiltonian $H^{\rm
el}(\{\Delta_q\})$ and corresponds to the LUMO (or HOMO) as modified by
the atomic distortion.  The minimisation of $E_{\rm
half}(\{\Delta_q\})$ is obtained when the forces
$F_q=M\omega_q^2\Delta_q+
\langle\varphi_e\vert\hat{\gamma}_q\vert\varphi_e\rangle
\sqrt{(2M\omega_q)/\hbar}$
are zero for all the phonon modes $q$, $\hat{\gamma}_q$ being the
$e$-ph coupling matrix with components $\gamma_{qnm}$.  With this
procedure, we can also study the contribution of the different phonon
modes $q$ and check the validity of using only a limited number of
optical phonon modes to create the distortions as we have already
suggested from the values of the Huang-Rhys factors.

Figure \ref{static_disto} shows the lattice distortions calculated within
different approximation from the ground state of a charged chain.
The lattice distortions are best represented by the staggered difference 
$d_i$ between adjacent bond lengths.
The quantity $d_i=(-1)^i(u_{i+1}-2u_i+u_{i-1})$ is known as the dimerization.
A constant dimerization pattern indicates a perfect bond length alternation
in the chain,
while a decrease of the dimerization indicates a deformation of the bond
lengths ({\it i.e.} an increase of the short bonds and a decrease of the
long bonds).
These lattice distortions are localized around the charge added and
are characteristic of the polaron defect in the molecular chain.
The general shape of the dimerization pattern in Fig. \ref{static_disto} 
indicates that a static polaron has formed in the chain and it extents over 
several atomic sites.

The dimerization patterns obtained from the original SSH model are shown
on Figure \ref{static_disto} with the dimerizations patterns
calculated by considering half the electronic spectrum and
classical phonon modes (all the modes and the reduced set of long 
wave-length optical modes).
We can see that the lattice distortions representing a static polaron 
defect in the chain are very well reproduced by considering only
the long wave-length optical phonons (for example those shown on Fig. 
\ref{fig_optmodes} for the $N_a=100$).
The discrepancies between the dimerization amplitudes obtained for
the full and half electronic spectrum are more important for long chains 
than for short chains.
For the short chains, the difference in dimerization amplitude does not
exceed $\approx$ 10\%.
In the extreme case of the shortest two-atom chain, working with
half the electronic spectrum gives the exact results.
For the long chains, the difference in dimerization amplitude increases
and is $\approx$ 17\% for the $N_a=100$ chain length.
We attribute the origin of these differences to the fact that the gap 
of the molecular wires decreases with the chain length.
However, as we show in the next section, an asymptotic regime is reached
for chain length $N_a\ge 100$, where the gap becomes independent of the 
chain length.
We therefore can assume that for longer chains, the difference in
dimerization amplitude should not increase further.

We also calculated the corresponding lattice distortions using quantum
phonons.  The calculations were done by determining the ground state
of the fully quantum electron/phonon Hamiltonian Eq. (\ref{Hwire})
where we introduced a cut-off for the number of possible phonon
excitations.  Each harmonic quantum phonon mode can contain only up to
$n_{\rm occ}^{\rm max}$ quanta.  Note again, that the calculations
were performed with the $n,m$ sums running only over half of the
electronic spectrum.  Then, for the sums running over the originally
empty conductance band states, the ground state $\vert\Psi_0\rangle$
corresponds to the situation where an extra electron has been added to
the chain.  The equivalent situation corresponding to removing an
electron is obtained by summing the electronic eigenstates only over
the originally occupied valence band states and considering among
these eigenstates of Eq. (\ref{Hwire}) that with the highest eigenvalue.

In practice, here we choose to calculate the situation corresponding to
a chain charged with one extra electron.
Due to charge-conjugation symmetry, the results for hole injection will
be identical.
Once the ground state $\vert\Psi_0\rangle$ of Eq. (\ref{Hwire}) is
obtained, we can calculate the quantum average 
$\langle\delta_q\rangle=\langle\Psi_0\vert\delta_q\vert\Psi_0\rangle$
for the mean displacement of the phonon mode $q$, where $\delta_q$ is
given by Eq. (\ref{phdisp}).
The atomic displacements $\langle u_i\rangle$ induced by charging
are obtained from 
$\langle u_i\rangle=\sum_q V_q(i)\ \langle\delta_q\rangle$.
The resulting dimerization patterns are shown on Figure \ref{static_disto}.
Convergence of the results is obtained for a small and finite number
of quanta in each mode, roughly $n_{\rm occ}^{\rm max}\approx 4,5,6$.
We will show in the next section that the results for the transport 
properties of the molecular wire coupled to the electrodes converge
faster with respect to the values of $n_{\rm occ}^{\rm max}$,
especially when one considers charge transport in the tunneling
regime. 
As expected, the dimerization patterns are very close to those obtained 
from the classical phonon model with half electronic spectrum.
The slight differences may be due to quantum delocalization of the 
eigenstate of Eq. (\ref{Hwire}).

Finally, it can be noticed that the general shape of the dimerization
induced by charging and the spatial extent of the corresponding 
polaron are well reproduced by the different approximations 
(half electronic spectrum, limited set of optical phonon modes, finite 
number of phonon excitations) introduced to reduce the Hilbert space.
The reduction of the parameter space is therefore entirely justified.
This reduction permits us to treat a great range of molecular
wire lengths and to determine the transport properties of the wires 
presented in the next section, with reasonable computing times.

\section{Results}
\label{results}

In this section, we present results for the electron transport
through typical heterojunctions made of a molecular wire connected
to two electron reservoirs.
In our calculations, the dynamical ({\it i.e.} energy-dependent) 
correlation between the electron and the phonon degrees of freedom is 
kept.
Such quantum coherence between the electron and the lattice is
essential to treat the propagation of polarons through the wire
in the tunneling regime ({\it i.e.} for an injection energy $E$ inside
the gap of the molecular wire).
We use the term coherence because there is no random dephasing of
the wave functions introduced by the electron-phonon interaction.
Instead each wave vector $\vert j,\{n_q\}\rangle$ has both a
definite amplitude and phase, which are both dependent on the
energy $E$ \cite{bonca97}.

\subsection{Lattice distortions induced by a tunneling electron}
\label{e_ind_disto}

In order to analyse the response of the molecular wire lattice
to a tunneling electron in the stationary state,
we calculate the expectation value of a correlation function 
between the phonon field displacements and the electron density.
Such a correlation function is defined as the following quantum 
average
\begin{equation} 
\delta_q^{[i]}=\frac
{\langle P_i\sqrt{\frac{\hbar}{2M\omega_q}}(a_q+a^\dag_q)P_i\rangle}
{\langle P_i\rangle}\ ,
\label{delta_q^i}
\end{equation} 
where $P_i=c^\dag_i c_i$ is the electron wave function projector
onto atomic site $i$ (the electron density operator on site $i$).
The quantum average of any electron and/or phonon operator $\mathcal O$
is given by 
$\langle{\mathcal O}\rangle=\langle{\mathcal O}(E)\rangle
=\langle\Psi(E)\vert{\mathcal O}\vert\Psi(E)\rangle$.
The correlation function $\delta_q^{[i]}$ represents the mean
displacement of the phonon mode $q$ when the electron is on site $i$.
We can then define the conditionally averaged atomic displacement 
$x_j^{[i]}=\sum_q V_q(j)\ \delta_q^{[i]}$
representing the atomic displacement on site $j$ on the condition that 
the electron is on site $i$.
From these atomic displacements, we can obtain the dimerization pattern 
$d_j^{[i]}=(-1)^j(x_{j+1}^{[i]}-2x_j^{[i]}+x_{j-1}^{[i]})$.

\begin{figure}
\psfig{figure=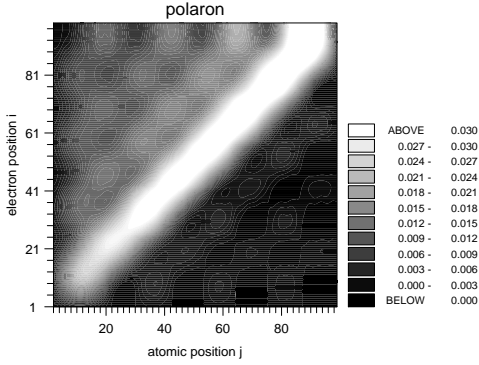,width=8cm}
\caption{
\label{dimer2D}
2D (atomic position - electron position) map of the dimerization pattern 
(in \AA)
obtained from the atomic displacements $x_j^{[i]}$ for a $N_a=100$ 
chain length.
For the boundary conditions chosen, an electron is injected from the left
(injection energy $E=0.0$ at mid-gap).
The charge current is flowing from atomic site $j=1$ to site $j=100$
(horizontal axis).
The bright parts represent a dip in the dimerization pattern, {\it i.e.}
the formation of a virtual polaron.
The induced atomic distortions are always located around the (tunneling)
electron position $i$ (vertical axis) when there is enough ``room'' for
the polaron to exist inside the chain.}
\end{figure}

\begin{figure}
\psfig{figure=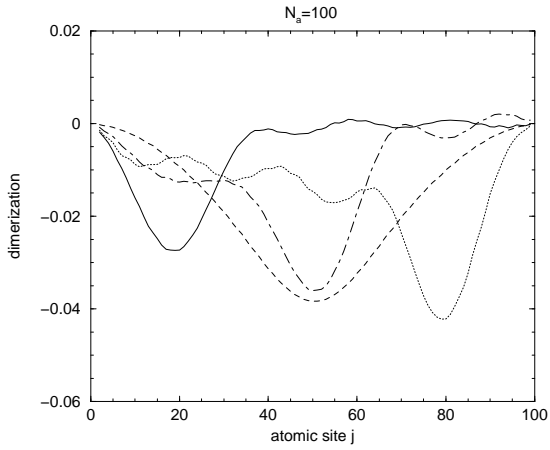,height=6cm}
\caption{
\label{dimer1D}
Dimerization pattern obtained from the atomic displacements $x_j^{[i]}$ for 
$N_a=100$ (identical to figure \ref{dimer2D}) for three different
electron positions $i=19$ (solid line), $i=51$ (dot-dashed line) and 
$i=81$ (dotted line).
The corresponding dimerization pattern for a static polaron inside an
isolated chain is also shown (thin dashed line, identical to the
dimerization shown in figure \ref{static_disto} as the dot-dashed
line).
Note the difference in the width of the polaron defect for a static
electron and a tunneling electron. 
Note also the ``wake'' of lattice distortion left by the tunneling
electron.}
\end{figure}

We have calculated the response of the molecular lattice to the tunneling
electron for different wire lengths and different injection energies $E$.
For example, Figure \ref{dimer2D} shows a two-dimensional map 
(atomic position $j$/electron position $i$) of the dimerization 
$d_j^{[i]}$ for a $N_a=100$ chain length.
The tunneling electron is injected at mid-gap ($E=0.0$) from the left
($j=1$) and propagates to the right ($j=100$).
The bright part around the first diagonal in Fig. \ref{dimer2D} represents 
a dip in the dimerization pattern along the atomic positions.
As shown in the previous section, such a lattice distortion is the signature 
of the formation of a polaron inside the chain.
We name it a virtual polaron because of the transient nature of
the tunneling electron injected inside the molecular wire gap. 
The induced atomic distortions are located around the electron 
position $i$ (vertical axis).
As can be seen in Figs. \ref{dimer2D} and \ref{dimer1D}, the virtual polaron 
has its own intrinsinc width (estimated around $\approx$ 15 atomic sites for
the present model of molecular chains which is consistent with the 
correlation length $\xi$ of the continuum model \cite{note1}).
Its formation is therefore not possible for the electron positions at the 
ends of the molecular wire.
For short chains ($N_a<15$), the lattice distortion associated with the 
polaron cannot be accommodated in the wire.
This has an important consequence on the electronic transport properties,
as we shall see in the next section.

The atomic displacements are, for most electron positions, slightly less 
than those for an isolated chain (Fig. \ref{dimer1D}) because the lattice 
does not respond fully to the tunneling electron as in the case of the 
static charge added into the chain.
The width of the lattice distortion around the electron is also 
smaller for the tunneling electron than for the static charge.

The amplitude of the virtual polaron slightly increases when the 
injection energy $E$ increases above mid-gap.
Such a behaviour persists until $E$ reaches the first resonance peak in 
the electron transmission (see next section).
This may be understood from the fact that far from mid-gap, the
electron wave function gets more weight (the amplitude of the
electron wave function gets larger) leading to a stronger coupling
to the phonons and therefore to more important lattice distortions. 
Above the first resonance peak, the electron is not, strictly speaking,
tunneling anymore through the molecular gap.
For these energies, the transport regime is better described as 
tunneling resonantly through the coupled $e$-ph states of the system.
The scattering wave function acquires more weight from these quasi-standing
waves which in turn modify qualitatively the polaron-like nature of
the corresponding atomic distortions.
A detailed analysis of such distortions is out of the scope of 
this paper and will be presented elsewhere \cite{ness_perturb}.

The coherent electron-lattice distortion leads to a modification of the 
electronic spectrum of the molecular wire compared to the spectrum of the 
undistorted chain. 
In all the cases studied, the polaron formation is associated with a 
reduction of the gap of the originally undistorted chain.
Because the atomic distortions induced in the case of a static electron 
added in the chain are different from those obtained for a tunneling 
electron, we expect the electron transport properties through the 
corresponding spectra to be different.
The behaviour of the transport properties in the molecular wires is shown 
and analysed in terms of electron transmission probabilities in the next 
section.

\subsection{Transport properties}\label{trans_prop}

The current flowing through the different channels (for example in the 
right outgoing channels) is given by
\begin{equation}
j^{\rm R}_{\{m_q\}}(E)=\frac{2e}{h}\ \Im m\ 
\left(
\alpha_{l,\{m_q\}}^*\ \beta_{\rm R}\ \alpha_{l+1,\{m_q\}}
\right)\ ,
\label{jR}
\end{equation}
for sites $l>N_a+1$ located inside the right lead.
From the asymptotic form of the wave functions inside the leads Eqs.
(\ref{left_wf}) and (\ref{right_wf}), it is found that 
$j^{\rm R}_{\{m_q\}}$ is related to the transmission probability
by $j^{\rm R}_{\{m_q\}}=2\frac{e}{h}\beta_{\rm R}\sin k^{\rm R}_{\{m_q\}}\ 
\vert t_{\{m_q\}}(E)\vert^2$.
Similarly, the current inside the left lead channels 
$j^{\rm L}_{\{m_q\}}$ is related to the reflection probability 
$\vert r_{\{m_q\}}(E)\vert^2$.

We can define an effective total transmission probability as
\begin{equation}
T(E)=\sum_{\{m_q\}}\vert t_{\{m_q\}}(E)\vert^2\ 
\frac{\beta_{\rm R}\ \sin k^{\rm R}_{\{m_q\}}}
{\beta_{\rm L}\ \sin k^{\rm L}_{\{0\}}} \ .
\label{totTE}
\end{equation}
This takes the form of a sum of contributions from the different
outgoing channels \cite{note2}.

In the following, we present results for the injection of an electron
and transport by tunneling effect inside the gap and by resonant 
tunneling through the levels of the system above the gap.
The electron injection energy is defined as positive with respect to 
the reference energy $E=0$ for all the molecular wires.
The Fermi energies of the leads are assumed to be pinned at mid-gap in 
absence of any applied bias.

\begin{figure}
\psfig{figure=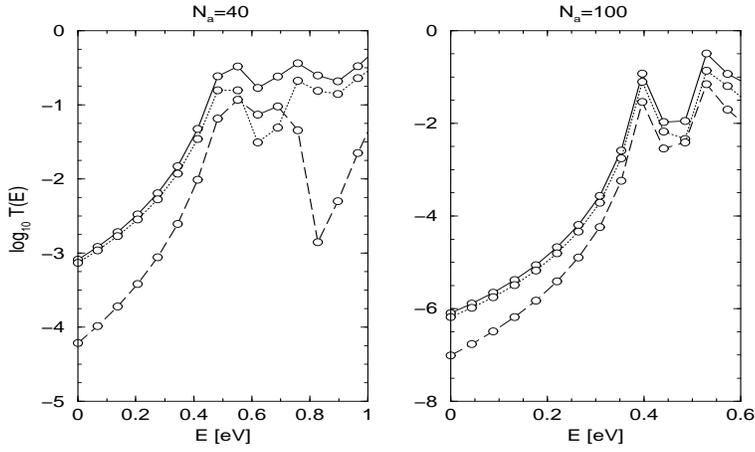,width=10cm,height=6cm}
\caption{
\label{trans_tunnel}
Electron transmission probability (on logarithmic scale) versus electron 
injection energy $E$ through molecular wires of different length.
Calculations were performed with $N_{\rm ph}=6,\ 
n_{\rm occ}^{\rm max}=2$ for $N_a=40$ and with 
$N_{\rm ph}=4,\ n_{\rm occ}^{\rm max}=3$ for $N_a=100$. 
Effective total transmission $T(E)$ (solid lines), transmission
from the elastic channel $\vert t_{\{0\}}(E)\vert^2$ only (dotted
lines), transmission from the first inelastic channel $\{n_{q=1}=1$
and $n_{q>1}=0\}$ (dashed lines). We recall that the 
first optical mode $q=1$ is the optical mode with the longest wave-length
({\it i.e.} lowest energy). The corresponding energy is 
$\hbar\omega_q$=0.148 eV (for $N_a=40$) and 
$\hbar\omega_q$=0.136 eV (for $N_a=100$).
In the tunneling regime, {\it i.e.} for energy E below $\approx$ 0.51 eV
(for $N_a=40$), and $\approx$ 0.39 eV (for $N_a=100$), the main contribution 
to $T(E)$ comes from the elastic channel.}
\end{figure}

Figure \ref{trans_tunnel} shows typical results for the effective
total transmission $T(E)$ and the contribution from the elastic
channel and from one inelastic channel.  The different transmission probabilities have, as
expected, an exponential behaviour in the tunneling regime (below the
first resonance peak) and present resonances through the spectrum of
the system for higher energies.
 
It is interesting to note that in the tunneling regime when the
injection energies are below the first resonance peak ($E<0.51$ eV for
$N_a=40$ and $E<0.39$ eV for $N_a=100$), the most important
contribution to the total transmission $T(E)$ comes from the elastic
channel.  Above the first resonance, the contribution of the other
inelastic channels may become as important as the contribution of the
elastic channel.  Although the contribution of the elastic process is
dominant in the tunneling regime, it should be noted that this
contribution is quite different to that obtained from an elastic
scattering treatment of the transport through the undistorted
molecule.  This is because the `elastic channel' corresponds to
processes in which, once the electron has crossed the molecule, there
is no overall absorption or emission of phonons.  Nevertheless,
phonons are emitted and absorbed during the intermediate stages of the
transport, and therefore this `elastic channel' is quite different
from a rigid-molecule calculation where no coupling to the phonons is
present.  We have already shown in Ref. \cite{ness99} that the
transmission is actually enhanced (in the tunneling regime and in the
limit of low temperatures) because of the $e$-ph coupling.  The
transmission enhancement comes from an effective gap reduction
associated with the virtual polaron formation.  This gap reduction is
also different from the one obtained by charging a (classical lattice)
chain with a static electron.

\begin{figure}
\psfig{figure=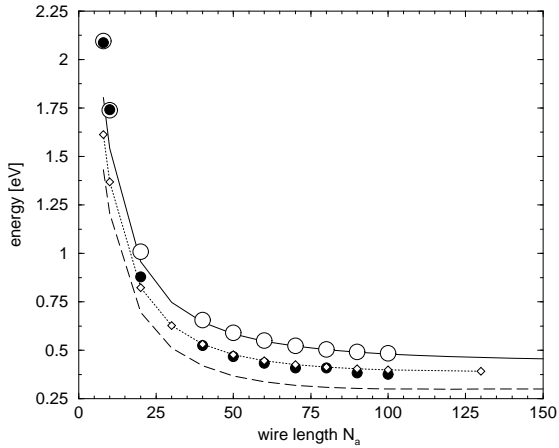,height=6cm}
\caption{
\label{gap_and_co}
Half gap obtained from the original SSH model versus the chain length
for a neutral molecule (solid line) and for a molecule charged with one 
extra electron (dashed line). 
The corresponding charging energy is also shown (dotted line with diamonds).
The energy positions of the first resonance peak in the transmission are 
also shown: for purely elastic scattering, {\it i.e.} ignoring the coupling 
to the phonons (empty circles), and for inelastic scattering including the 
coupling of the electron to the $N_{\rm ph}=4$ lowest frequency optical 
modes (with $n_{\rm occ}^{\rm max}=2$) (filled circles).
All energies are given in eV.}
\end{figure}

Figure \ref{gap_and_co} represents half the HOMO-LUMO gap of isolated 
neutral and charged chains obtained from the SSH model and the effective 
gap of our coupled quantum electron-phonon model for different chain 
lengths.
Half the gap corresponds to the value of the lowest unoccupied
molecular states (LUMO) of the isolated (neutral or charged) chain.
The effective gap of the molecular wire (including $e$-ph coupling)
connected to the leads is obtained from the position of the first 
resonance peak in the transmission $T(E)$.
We also plotted the position of the first resonance peak obtained from
purely elastic calculations (ignoring the $e$-ph coupling).
As expected, the positions of these resonances reproduce the values of
half the gap of the corresponding neutral chains.
An energy shift appears in the first resonance position because
of the real part of the embedding potentials $\Sigma^{\rm L}$ and
$\Sigma^{\rm R}$ that appear in the solution of the scattering problem
Eq. (\ref{linsys}).

For the fully inelastic scattering calculations, the first resonance in 
$T(E)$ occurs for injection energy $E$ smaller than those obtained from
the (solely) elastic calculations.
The behaviour characterises the effective gap reduction of the wire
due to the $e$-ph interaction.
The energy position of the first resonance in $T(E)$ is close to the 
charging energy $E^{\rm charg}$ of the isolated molecular wire.
The charging energy $E^{\rm charg}$ is defined as the difference
between the (selfconsistent) ground state total energy of the charged 
chain $E_{+1}[u_i^{\rm c}]$ (with the distorted lattice positions
$u_i^{\rm c}$ corresponding to a static polaron)
and the ground state total energy of the neutral chain 
$E_0[u_i^0]$ (with the undistorted, perfectly dimerized, lattice
positions $u_i^0$).
Some values of $E^{\rm charg}$ for different molecular wire lengths are 
given in Table \ref{SQandEner}.
Differences between the values of $E^{\rm charg}$ and the first resonance
in $T(E)$ do occur, they are due to (i) the systematic energy shift 
introduced by the embedding potentials and also to (ii) the differences 
arising from treating the lattice distortions classically or with quantum 
phonons as pointed out in section \ref{e_ind_disto}.

To summarise, there are two distinct physical processes that affect the
value of the band-gap of the molecular wire and therefore the transport
properties through this wire.
Firstly, the intrinsic band-gap dimishes with increasing wire lengths.
The values reach a asymptotic regime for a length $N_a\stackrel{>}{\sim}100$.
The asymptotic band-gap value and the length above which the asympotic
regime is obtained depend of the chemical nature of the molecule,
{\it i.e.} on the SSH parameters used to model the molecular chain
(cf. Appendix \ref{appA}).
Secondly, an effective band-gap reduction occurs upon charging the 
molecular wire with a static charge or with a transient tunneling
charge.
The reduction is due to the the coupling between the charge and
the lattice leading to the formation of static or virtual polaron
respectively.
Therefore, the electron transmission through the molecular wire
increases because of the $e$-ph coupling for injection energies
inside the gap.
Although, the trends for the gap reduction are similar for the
static and virtual polarons, they differ quantitatively over the whole 
range of molecular lengths studied here.
   
Finally, it should be noticed that for short wires ($N_a\le 10$) the 
effective gap obtained from the fully inelastic calculations converges
towards the gap obtained from elastic calculations.
Therefore the values of the electron transmission are almost identical
for both inelastic and purely elastic calculations.
As mentioned in section \ref{e_ind_disto}, although the polaron cannot
be accomodated in very short wires, in these conditions
(short tunneling length and large gap), the tunneling process
itself is too fast to get significant lattice distortions associated with
the charge injection. 
In order to illustrate this point, we give in Table \ref{charac_time}
the characteristic times associated with the electron and the phonons.
We calculate the time domain $\tau_{\rm ph}$ of the phonons from the
extremal phonon frequencies used in our calculations.
We estimate a residence time $\tau_{\rm res}$ for the electron from
the full width at half maximum $\eta$ of the first resonance peak using 
$\tau_{\rm res}=\hbar/2\eta$.
Following B\"uttiker and Landauer \cite{landauer94,buttiker85-82}, 
a traversal time $\tau_{\rm BL}$ for 
the tunneling electron can be obtained; it is calculated from the elastic 
channel transmission coefficient as done in Ref. \cite{ness98}.
The traversal time $\tau_{\rm BL}$ is by definition dependent on the electron
injection energy $E$.
We give in Table \ref{charac_time} the range of $\tau_{\rm BL}$ corresponding
to energies inside the tunneling gap. 
We can see that for short wires ($N_a \ll 2\xi$), the tunneling time
(and also $\tau_{\rm res}$) is smaller than the characteristic times 
associated with the phonons.
The lattice dynamics is therefore not fast enough to respond to the tunneling
electron.
In this regime, the transport properties obtained from purely elastic scattering
(rigid lattice) are not strongly different from those obtained by inelastic
scattering (coherently distorted lattice by the $e$-ph coupling).
For longer wires ($N_a\approx 40\gg 2\xi$), the tunneling time
(and $\tau_{\rm res}$) is comparable to or larger than the characteristic
phonon times; the polaron can be formed inside the molecular wire.
It is in this regime that we observe the most important differences
between the electron transmission for a rigid lattice, and for a lattice
that can be deformed by the tunneling electron.

\subsection{Lattice fluctuations}\label{latt_fluctu}

The importance of lattice fluctuations on the electronic structure 
of conjugated molecules has already been considered 
\cite{mckenzie92,takahashi92,galli95,yu97}.
As the lattice fluctuations (even in the limit of zero temperature, 
{\it i.e.} the zero-point motion) are of the same order
of magnitude than the distortions induced by charge injection, it is
important to know if such species (polarons, solitons) survive the 
lattice fluctuations.

We present here results for the electron transmission through a molecular 
chain where disorder is introduced due to lattice fluctuations.
In order to compare these results with the full quantum inelastic 
transmission, we consider the limit of low temperatures where the
lattice fluctuations are due to the zero-point motion of each
phonon mode.
The averaged transmission is obtained by summing up the different
elastic transmission probabilities $T(E;\{\Delta_q\})$ associated with 
the displaced phonon configurations $\{\Delta_q\}$, weighted by the 
Gaussian distribution probability of the ground state harmonic oscillator 
wave function. 
The average of a function $f(\{\Delta_q\})$ depending on the phonon
displacements $\{\Delta_q\}$ is calculated as
\begin{eqnarray}
\langle f\rangle_{\rm dis} = 
\int d\Delta_1 ... d\Delta_q ... d\Delta_{N_{\rm ph}} \ 
f(\{\Delta_q\})
\prod_q\frac{\exp(-\frac{1}{\hbar}M\omega_q \Delta_q^2)}
{\sqrt{2\pi}\sigma_q} \ ,
\label{avfunc}
\end{eqnarray}
where the width of the Gaussian distribution is given by the virial 
theorem for the zero-point fluctuation of each mode 
$
\frac{1}{2}M\omega_q^2\langle X_q^2\rangle=\frac{1}{2}E_{n=0}=
\frac{1}{2}(n+\frac{1}{2})\hbar\omega_q=\frac{1}{4}\hbar\omega_q
$
and $\sigma_q=\langle X_q^2\rangle^{1/2}$.

When $f(\{\Delta_q\})=T(E;\{\Delta_q\})$, Eq. (\ref{avfunc}) is equivalent 
to the static lattice approximation expression obtained by Pazy and 
Laikhtman \cite{pazy99} using a path-integral formalism.
In our calculations, the transmission $T(E;\{\Delta_q\})$ is calculated from 
elastic scattering through the one-electron spectrum of the molecular chain 
distorted by the zero-point lattice fluctuations. 
The distorted molecule eigenstates are obtained from the Hamiltonian 
Eq. (\ref{SSHamilt}) by replacing the equilibrium atomic positions $u_i^0$ 
by $u_i=u^0_i+\delta u_i$ where $\delta u_i=\sum_q V_q(i) \Delta_q$ for 
the different configurations $\{\Delta_q\}$ of the distorted lattice.
In practice, $T(E;\{\Delta_q\})$ is calculated in a similar way as 
described in section \ref{MCST} but using the approximation of elastic 
scattering ({\it i.e.} no energy exchange between the electron and
the phonons is allowed, setting $n_{\rm occ}^{\rm max}=0$ for all the 
phonon modes considered).
Futhermore, the integrals over the phonon displacements are performed
using an algorithm to generate random deviates with a normal Gaussian
distribution with a given zero mean value and a variance $\sigma_q$.

The transmission Eq. (\ref{avfunc}) corresponds to an average over the
phonon modes, treated classically (as in Appendix \ref{harmonic_ph})
but with a mean-square displacement equal to the quantum zero-point
motion, while the electronic transmission $T(E;\{\Delta_q\})$ is
obtained from quantum mechanics.  In this average the influence of the
phonon displacements on the electronic spectrum is taken into account
but not {\it vice versa}.  This is the fundamental difference with the
transport calculations presented in section \ref{trans_prop} where
both the electron and phonon degrees of freedom are treated at the
quantum level and where the lattice distortion is induced by the
injected electron.  We therefore expect the results for zero-point
motion to be different from the one due to the quantum coherent
electron-phonon coupling.

\begin{figure}
\psfig{figure=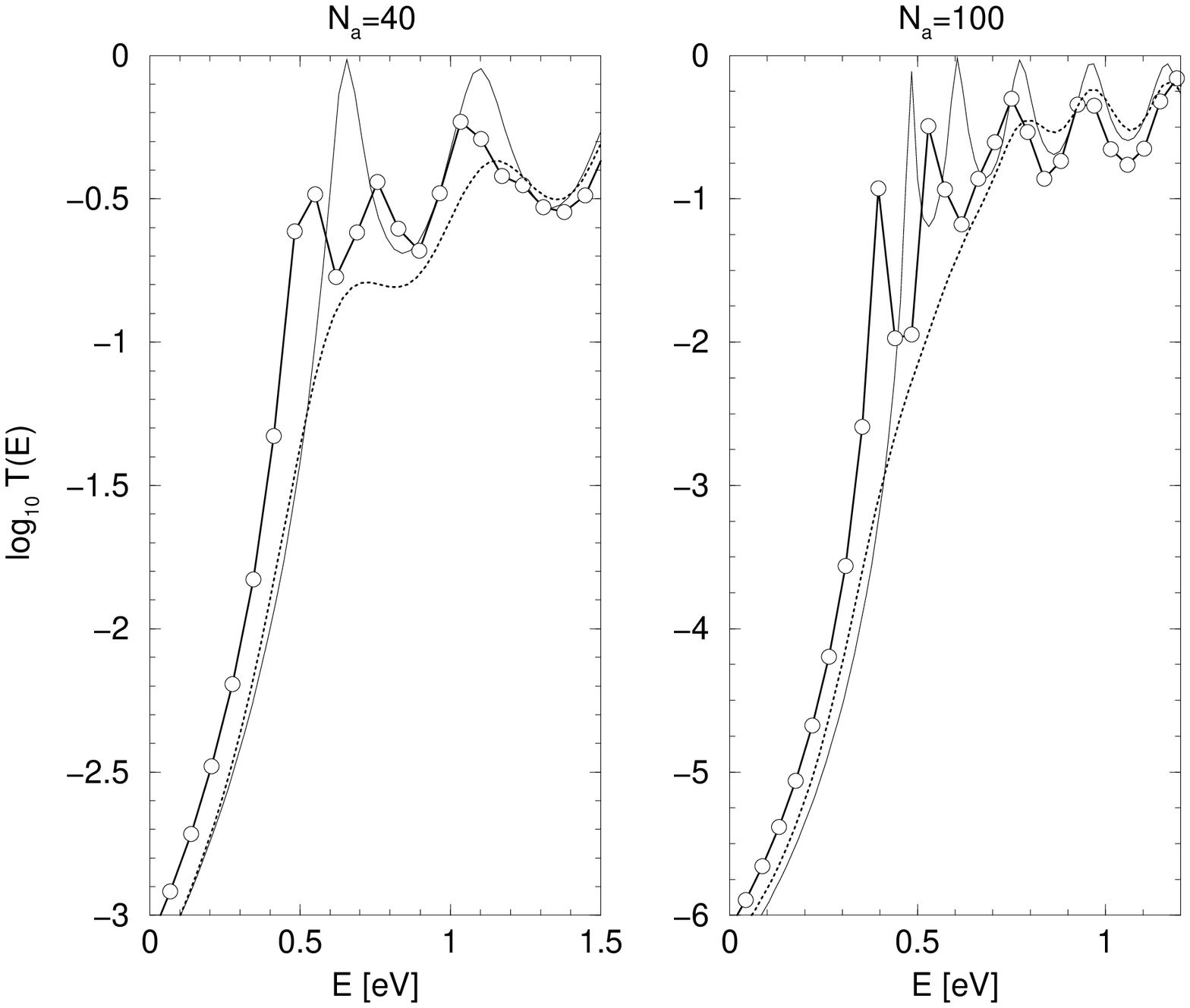,width=9cm,height=6cm}
\caption{
\label{av_trans_and_co}
Electron transmission $T(E)$ versus electron injection
energy $E$ for two molecular wire lengths $N_a=40$ and $N_a=100$.
Transmission through the one-electron eigenstates of the neutral chains
from elastic scattering (thin solid lines).
Transmission through the coupled quantum electron/phonons system from
inelastic scattering (solid lines with circles), 
$N_{\rm ph}=6$ and $n_{\rm occ}^{\rm max}=2$ for $N_a=40$, 
$N_{\rm ph}=4$ and $n_{\rm occ}^{\rm max}=3$ for $N_a=100$.
Averaged transmission $\langle \log T(E)\rangle_{\rm dis}$ from elastic 
scattering through the one-electron eigenstates of the chain
distorted by zero-point lattice fluctuations (dotted lines),
$N_{\rm ph}=3$ for $N_a=40$ and $N_a=100$.}
\end{figure}

Figure \ref{av_trans_and_co} shows the elastic transmission probability 
through the rigid undistorted lattice and the transmission averaged (in 
the manner discussed below) over the zero-point fluctuations.
The total effective transmission obtained from the inelastic scattering
calculations is also shown.

As expected, the lattice fluctuations substantially modify the electron 
transmission through the molecular wire.
The transmission is reduced around the resonance peaks of the one-electron 
spectrum of the undistorted chain, and the peaks are smeared out.
However, there is no shift in the position of the resonances 
corresponding to the formation of virtual polarons. 

As in the fully quantum calculations, the transmission is enhanced for 
injection energies inside the gap of the undistorted chain.  
However, at this point it becomes important to know exactly how the 
average over the lattice fluctuations in Eq.~(\ref{avfunc}) is 
calculated.  
It is known \cite{markos93} that the ensemble average of the transmission 
probability in a disordered one-dimensional conductor is a statistically 
ill-defined quantity, in the sense that it is dominated by a very small 
number of exceptional configurations.  
In other words, the mean is in no way representative of the typical 
transmission probability to be expected from a randomly sampled member 
of the ensemble.  
This difficulty does not arise if $\log T$, rather then $T$, is 
averaged \cite{kirkman84}.
In Figure~\ref{av_trans_and_co} we therefore compare $\log T$ from our 
fully quantum results with  $\langle\log T\rangle_{\rm dis}$.
In this case we see that fully including the electron-phonon coupling
(allowing the possibility of polaron formation) gives a greater increase 
in the tunneling transmission than averaging over the disorder 
introduced by the zero-point motion.
This shows that only the fully quantum coherent electron-phonon coupling 
calculations (exact in the limit of zero temperature) give the correct 
physics.

\section{Conclusion}\label{ccl}

In this paper, we have presented a method to calculate the inelastic
effects on electron transport through one-dimensional molecular wires, 
including realistic electron-phonon coupling.
The model for the wires is inspired by the Su-Schrieffer-Heeger
model for trans-polyacetylene.
The transport through the quantum electron-phonon system is solved
by means of a multichannel scattering technique where each channel
is associated with probabilities for the electron to be reflected or
transmitted, given the phonon occupation number configuration.

The results show that the transport in the one-dimensional molecular 
wires does not occur as in traditional (three-dimensional) 
semiconducting molecular devices.
It does not involve the propagation of free electron-like particles, 
but instead is due to the coherent propagation of ``quasi-particles''~:
an electron (or hole) surrounded by a lattice distortion, 
characteristic of the formation of an electron (or hole) polaron.
This object is called here a virtual polaron because of the transient
nature of the tunneling electron (or hole) injected inside the 
HOMO-LUMO gap of the molecular wire.
In this regime and in the limit of low temperatures, the tunneling
transmission probability through the device increases, due to the 
electron-phonon coupling, in comparison with the transmission obtained 
from elastic scattering through the undistorted molecular wire.
Lattice fluctuations also modify the electron transmission through the
wire.
However, the corresponding enhancement of the transmission in the 
tunneling regime is less than that produced by the virtual polarons.

The influence of other defects (such as a soliton-like defect existing 
in the chain) on the transmission has already been considered, and will 
be presented elsewhere.
The transport for finite temperatures (different initial phonon 
occupations and proper statistical averages) is under study and results
will be presented in the near future.

The results presented in this paper are general and can be applied to
other types of one-dimensional atomic-scale wires subject to a Peierls 
transition.
For example, it has been shown both experimentally and theoretically
that a Peierls-like transition also occurs in dangling-bond (DB) lines 
fabricated on the H-passivated Si(001) surface 
\cite{hitosugi99,hitosugi98,doumergue99}.
More recently, it has been shown theoretically that the injection of
a static charge in the bands around the band-gap leads to a distortion 
of the atomic positions along the line \cite{bowler00}.
This distortion corresponds to the formation of a small polaron in the 
DB line.
We therefore expect that carrier injection in the band-gap of the DB 
lines will lead to similar physical results to those presented here 
for molecular wires.
Many other quasi-one-dimensional systems may be expected to show 
similar characteristics.

\acknowledgments

The authors greatly appreciate enlightening discussions with 
L. Kantorovich, J. Gavartin, A.L. Shluger and A.M. Stoneham.
HN thanks C. Joachim for stimulating discussions on inelastic tunneling
in molecular systems.
We acknowledge support from the U.K. Engineering and Physical Sciences 
Research Council in the form of an Advanced Fellowship (AJF),
a Postdoctoral Fellowship (HN), a Research Studentship (SAS) and under 
Grant No. GR/M09193.

\appendix

\section{Isolated molecular wires}\label{appA}
\subsection{The SSH model}

Su, Schrieffer and Heeger (SSH) modelled a chain of trans-polyacetylene
($t$-PA) as a purely one-dimensional atomic (CH)$_x$ chain~\cite{SSH88,SSH}.
The model combines a classical ball-and-spring term for the distortion
of the $\sigma$-bond backbone with a tight-binding representation for
the delocalised $\pi$-electron orbitals along the chain. Furthermore
the electron hopping integrals between adjacent (CH) groups are
expanded linearly about some reference values. The corresponding
Hamiltonian is 
\begin{equation}
H_{\rm SSH}=-\sum_{i,s}\left(t_0-\alpha\left(u_{i+1}-u_i\right)\right)
\left[c^\dag_{i,s}c_{i+1,s}+c^\dag_{i+1,s}c_{i,s}\right]
+
\frac{1}{2} K \sum_i\left(u_{i+1}-u_i\right)^2 \ ,
\label{SSHamilt}
\end{equation}
where $c^\dag_{i,s}$ ($c_{i,s}$) creates (annihilates) a $\pi$-electron of
spin $s$ at site $i$. $u_i$ is the displacement of the $i$-th (CH) group
from its place in the reference (undimerised) system with hopping integral
$t_0$. $K$ is the spring constant corresponding to the $\sigma$-bond and
$\alpha$ is the electron-lattice coupling constant.
As explained in the introduction, the undimerised metallic chain is unstable 
with respect to a Peierls distortion, and the ground state of an infinite 
neutral chain has displacements given by $u_i=(-1)^i\ u_0$, where $u_0$ is
a constant depending on $t_0$, $\alpha$ and $K$. 
The values chosen for the different parameters ($t_0$=2.5 eV, 
$\alpha$=6.1 eV/\AA~and $K$=42.0 eV/\AA$^2$)~\cite{wallace89} give for an 
infinite perfectly dimerised chain a total band width of 10 eV, a band gap 
of 1.4 eV, and a difference between long/short bond lengths of 0.1 \AA~in 
agreement with experiments.

For finite size chains containing $N_a$ atomic sites (i.e. (CH) groups),
the ground state GS($N_a$,$N_e$) of Hamiltonian (\ref{SSHamilt}) can be 
obtained for neutral ($N_e=N_a$) or charged ($N_e\neq N_a$) chains, 
$N_e$ being the total number of $\pi$-electrons inside the chain.
The ground state is obtained when the restoring forces of the springs balance 
the electronic forces by solving~\cite{wallace89,stafstrom84,chao85,xie93}
\begin{equation}
2\alpha\sum_{n}^{\rm occ}Z_{n}(i)\left(Z_{n}(i-1)-Z_{n}(i+1)\right)
+
K\left(2u_i-u_{i-1}-u_{i+1}\right) = 0 \ ,
\label{GSequi}
\end{equation}
where $Z_n(i)$ are the components of the one-electron eigenstates
(with energy $\epsilon_n$) of $H_{\rm SSH}$ for a given atomic 
configuration $\{u_i\}$. 
The sum in Eq.(\ref{GSequi}) runs only over the occupied states and the 
spin index $s$ is implicitly taken into account in the $n$ summations.
For the finite size molecular chains we study here, the boundary 
conditions are $Z_n(i)=0$ and $u_i=0$ when $i<1$ or $i>N_a$.
Furthermore in practice, in order to avoid the uniform translation of 
the chain through space, one atomic site is kept fixed, for instance one 
end of the molecular chain is kept fixed (i.e. $u_{N_a}=0$).

\subsection{Classical phonons}\label{harmonic_ph}

In the limit of small displacements $\{\delta u_i\}$ around the equilibrium
atomic positions $\{u^0_i\}$ of the ground state GS($N_a$,$N_e$), it is possible 
to derive an effective harmonic phonon Hamiltonian by partitioning 
Eq.(\ref{SSHamilt}) as follows \cite{wallace89,chao85,xie93,sun87}
\begin{equation}
H_{\rm SSH}[\{c_n\},\{u^0_i+\delta u_i\}]=
H_{\rm st}[\{c_n\},\{u^0_i\}]+
H_{\rm ph}[\{u^0_i\},\{\delta u_i\}]+
H_{e-{\rm ph}}[\{c_n\},\{\delta u_i\}] \ ,
\label{expSSH}
\end{equation}
where $H_{\rm st}$, $H_{e-{\rm ph}}$, $H_{\rm ph}$ are respectively the static, 
phonon and electron-phonon coupling parts of the total SSH Hamiltonian.
$H_{e-{\rm ph}}$ is usually treated as a perturbation up to second order to
give a quadratic term in $\delta u_i$ in the effective phonon Hamiltonian
$\sum_{i,j} K_{ij}\ \delta u_i\delta u_j$. The dynamical matrix $K_{ij}$
is given by \cite{chao85}
\begin{equation}
K_{ij}=K_{ji}=2\alpha^2\sum_n^{\rm unocc}\sum_m^{\rm occ}
\left(
F(i,n,m)-F(i+1,n,m)
\right)
\left(
F(j,n,m)-F(j+1,n,m)
\right)
/
\left(
\epsilon_m-\epsilon_n
\right) \ ,
\label{dynmat}
\end{equation}  
where $F(j,n,m)=Z_n(j)Z_m(j-1)+Z_m(j)Z_n(j-1)$.
The $n$ ($m$) sum runs over the empty (occupied) electronic states.
The eigenstates of dynamical matrix $K_{ij}$ give the eigenmodes
$V_q(i)$ of vibration (phonons) of the finite size molecular chain,
while the eigenvalues, $M\omega_q^2$, of $K_{ij}$ are related to the
phonon frequencies $\omega_q$ ($M$ being the mass of the CH group).

In the present paper, we use the same boundary conditions as above to
determine the phonon modes. However, different boundary conditions
could be used : fixed ends (i.e. constant molecular chain length),
free ends eventually coupled to different spring constants to simulate
the effective coupling to the electrodes. 
Although, these different boundary conditions would in principle affect 
the electronic and the vibrational properties of the chain 
\cite{mikrajuddin00}, 
we infer that the main physical results obtained in the present study 
will not be drastically modified. 
For instance, it appears that these different conditions will mostly 
affect the acoustic modes of the chain. 
Those modes have been however neglected in the present work because their 
contribution to the (virtual) polaron formation is negligible.

\subsection{Quantum phonons}\label{QM_ph}

At this stage, we already have all the ingredients to derive a
quantum version of the SSH Hamiltonian.
From the reference system, chosen to be the neutral molecular
chain of length $N_a$, we can write the Hamiltonians for the non-interacting
electron and phonon degrees of freedom as
\begin{equation}
H_e=\sum_n \epsilon_n \ c^\dag_n c_n \ ,
\label{Helec}
\end{equation}
where $c^\dag_n=\sum_i Z_n(i)\ c^\dag_i$ creates ($c_n$ annihilates) an electron
in the $n$th electronic state of the reference system with energy $\epsilon_n$.
The harmonic phonon Hamiltonian (neglecting the zero-point energy) is 
\begin{equation}
H_{\rm ph}=\sum_q \hbar\omega_q \ a^\dag_q a_q \ ,
\label{Hphon}
\end{equation}
where $a^\dag_q$ ($a_q$) creates (annihilates) a phonon mode $q$ with 
frequency $\omega_q$.
The basis set associated to $H_e+H_{\rm ph}$ is formed by the eigenstates
$\vert n,\{n_q\}\rangle=c^\dag_n\prod_q \frac{(a^\dag_q)^{n_q}}{\sqrt{n_q!}}
\vert 0\rangle$
with eigenvalues $\epsilon_{n,\{n_q\}}=\epsilon_n+\sum_q n_q\ \hbar\omega_q$,  
where $\vert 0\rangle$ is the vacuum state and $\{n_q\}$ the set of phonon 
occupation numbers.
 
We expand the lattice deformations $\delta x_i$ induced by 
an additional charge introduced in the chain onto the phonon
modes of the neutral chain: $\delta u_i=\sum_q V_q(i)\ \delta_q$.
The new lattice positions, displaced from the equilibrium position $u^0_i$,
are $u_i=u^0_i+\delta u_i$. 
Then, the linear electron-phonon coupling term of the original SSH Hamiltonian 
is written in a quantum form by quantizing the phonon field displacements 
\begin{equation}
\delta_q=\sqrt{\frac{\hbar}{2M\omega_q}}
\left(
a_q+a^\dag_q
\right) \ .
\label{phdisp}
\end{equation}
Therefore the $e$-ph coupling Hamiltonian is
\begin{equation}
H_{e-{\rm ph}}=\sum_{q,n,m}
\gamma_{qnm}
\left(
a^\dag_q+a_q
\right)
c^\dag_n c_m \ ,
\label{Heph}
\end{equation}
where
\begin{equation}
\gamma_{qnm}=\sum_{i=2}^{N_a} \lambda_q(i)
\left[
Z_n(i)Z_m(i-1)+Z_n(i-1)Z_m(i)
\right] \ ,
\label{gamqnm}
\end{equation}
and 
\begin{equation}
\lambda_q(i)=\alpha[V_q(i)-V_q(i-1)]\times\sqrt{\frac{\hbar}{2M\omega_q}} \ .
\label{lamdaiq}
\end{equation}
The total Hamiltonian $H_{\rm w}$ for the molecular wire with quantum 
phonons and linear electron-phonon coupling inspired by the SSH model is 
given by the sum $H_{\rm w}=H_e+H_{\rm ph}+H_{e-{\rm ph}}$ as in 
Eq. (\ref{Hwire}).

Finally it should be noted that the $e$-ph coupling matrix elements 
$\gamma_{qnm}$ obey some selection rules.
Generally, $\gamma_{qnm}=0$ unless the direct product
$\Gamma_q\otimes\Gamma_n\otimes\Gamma_m$ contains the identity 
representation ($\Gamma_n$, $\Gamma_m$ and $\Gamma_q$ being the
irreducible representation of the eigenstate $n$, $m$ and of 
$\lambda_q$ respectively).
In practice, $Z_n$ and $V_q$ are even/odd functions with respect to 
the center of the molecule. 
The quantity $Z_n(i)Z_m(i-1)+Z_n(i-1)Z_m(i)$ is even (odd) when $n+m$ is 
an even (odd) integer (indexing the eigenvectors $Z_n$ by increasing 
eigenvalues and $Z_{n=1}$ being even).
Whenever the quantity under the site $i$ summation in Eq. (\ref{gamqnm}) 
is odd, $\gamma_{qnm}=0$. 
Although it is not surprising to obtain selection rules for the 
$e$-ph coupling, their existence is very important in order to 
reduce the computing time of the product $H_{\rm w}\vert\phi\rangle$
needed to solve Eq. (\ref{linsys}).

\newpage

\begin{table}
\caption{Huang-Rhys factors $S_q$, the
deformation energy $\Delta E=E_0[u_i^{\rm c}]-E_0[u_i^0]$,
the harmonic distortion 
energy $\sum_q \hbar\omega_q S_q$ and 
the charging energy $E^{\rm charg}=E_{+1}[u_i^{\rm c}]-E_0[u_i^0]$
for different molecular wire lengths. See main text for the
definition of the different quantities.}
\begin{tabular}{cccccc}
  & $N_a$=20 & $N_a$=40 & $N_a$=60 & $N_a$=80 & $N_a$=100 \\ \hline
largest $S_q$ & 0.806 & 0.855 & 0.901 & 0.885 & 0.825 \\
second $S_q$  & 1.03$\times$10$^{-3}$ & 6.13$\times$10$^{-4}$ & 
5.01$\times$10$^{-3}$ & 1.67$\times$10$^{-2}$ & 
3.66$\times$10$^{-2}$\\ \hline
$\Delta E$ [eV] & 0.127 & 0.111 & 0.108 & 0.104 & 0.098\\
$\sum_q \hbar\omega_q S_q$ [eV] & 0.136 & 0.127 & 0.128 & 0.125 & 0.118\\ \hline
$E^{\rm charg}$ [eV] & 0.823 & 0.531 & 0.446 & 0.413 & 0.399 
\end{tabular}
\label{SQandEner}
\end{table}

\begin{table}
\caption{Characteristic times associated to the electron and phonons.
$\tau_{\rm res}$ is the residence time for the electron estimated
from the width of the first resonance peak.
$\tau_{\rm BL}$ is the B\"uttiker-Landauer traversal tunneling time for 
the electron.
The values for $\tau_{\rm ph}$ are a range corresponding to the range
of optic phonon frequencies.
The different $\tau$ are given in fs.}
\begin{tabular}{ccccccccccc}
$N_a$& 08 & 10& 20& 40& 50& 60& 70& 80& 90& 100 \\ \hline
$\tau_{\rm res}$& 0.40& 0.51& 1.56& 5.51& 10.82& 15.05& 20.67& 30.21& 
37.5& 45.29\\
$\tau_{\rm BL}$&-& 0.77-0.90& 1.64-1.70& 3.35-4.50&-&-&-&-&-& 8.36-25.53\\
$\tau_{\rm ph}$& 3.34-3.88& 3.34-3.60& 3.36-3.93& 3.65-4.43& 
3.79-4.58& 3.78-4.68& 4.04-4.74& 3.99-4.79& 4.23-4.82& 4.31-4.84
\end{tabular}
\label{charac_time}
\end{table}

\begin{thebibliography}{99}
\bibitem{note_tun_inel}
In a different context (atomic manipulation with a scanning tunneling tip), 
a model for studying tunneling current heating of atomic adsorbates has been 
recently developed \cite{joachim_new}.
This model, valid for low energy vibrational state excitations, goes
beyond the wave packet approach to tunneling.
\bibitem{note1}
For the parameter used in this study, the gap parameter for an infinite
chain is given by $2\Delta=8\alpha u_0=1.4{\rm \ eV}$ (with a band width
$W=4t_0=10{\rm \ eV}$).
The correlation length is $\xi=W/\Delta\approx 7$ lattice parameters.
From the continuum model \cite{SSH88,TLM80}, the polaron width is 
$\approx 2\xi$ which is consistent with the results found here for the
finite length, discrete lattice, chain.
\bibitem{note2}
For the initial conditions used in the present study, the transmission
coefficients $t_{\{n_q\}}$ can be related to the inelastic transmission
coefficients $t(E_{\rm in},E_{\rm out})$ where $E_{\rm in}$ is the
energy of the injected electron and $E_{\rm out}$ is the energy of the 
outgoing electron in a given channel $\{n_q\}$. The relation between
the coefficients is $t(E_{\rm in},E_{\rm out})=\sum_{\{n_q\}}
t_{\{n_q\}}(E_{\rm in})\ \delta(E_{\rm in}-E_{\rm out}-
\sum_q n_q\ \hbar\omega_q)$.
\end{thebibliography}
\end{document}